\documentclass{ws-ijgmmp}

\begin{document}

\markboth{Raymond STORA} {Renormalized perturbation theory}

%
\catchline{}{}{}{}{}
%

\title{RENORMALIZED PERTURBATION THEORY:\\
 A MISSING CHAPTER}

\author{RAYMOND STORA}

\address{CERN, Geneva, Switzerland \footnote{CERN, Theory Division, 1211 Gen\`eve 23} \\ and \\ LAPTH, Universit\'e de Savoie,
CNRS,\\Annecy-le-Vieux, France \footnote{Laboratoire
d'Annecy-le-Vieux de Physique Th\'eorique, UMR5108, 9 chemin de
Bellevue, BP 110, F-74941 ANNECY-LE-VIEUX Cedex, France}\\}



\maketitle

\begin{history}
\received{(Day Month Year)} \revised{(Day Month Year)}
\end{history}

\begin{abstract}

Renormalized perturbation theory \`a la  BPHZ can be founded on
causality as analyzed by H. Epstein and V. Glaser in the
seventies.

Here, we list and discuss a number of additional constraints of
algebraic character some of which have to be considered as parts
of the core of the BPHZ framework.

\end{abstract}


\section{Introduction}

Why a talk on Renormalized Perturbation Theory (RPT) in 2008? The
consensus established in the 70's under the acronym
BPHZ\footnote{BPHZ: Bogoliubov, Parasiuk, Hepp, Zimmermann, see
\cite{Bo-Sh}, \cite{He}, \cite{Z}, \cite{Ste}.} is part of
elementary particle physicists' theoretical equipment.

Yet, the corresponding literature is hard to penetrate for a mind
endowned with good logical connections - typically that of a
professional mathematician. This state of affairs may be assigned,
in parts, to some fuzziness about the connection between the
operator version and the functional version (\`a la Feynman) of
quantum mechanics, in this context.

In standard textbooks\cite{Wei}, the latter are both usually
described in formal terms which are most of the time not subject
to any mathematical formalization \footnote{e.g. equal time
commutation relations for interacting fields, writing down
sharp-time time ordered products.}.

Parallel to the establishment of BPHZ, and following the path
indicated by E.C.G. Stueckelberg\cite{Stu-Pe}\cite{Stu-Ri}, N.N.
Bogoliubov\cite{Bo-Sh} and coworkers, H. Epstein and V.~Glaser
\cite{EG} have opened the road -also in the 70's- to such a
mathematization. This path has been scarcely followed for the main
following two reasons, as it seems:

\begin{itemize}
\item BPHZ has been found tight enough within the physics
community mostly concerned with a large variety of interesting
topics. \item E.G.'s constructions have remained beyond a fence
which keeps their work isolated form concrete application of BPHZ,
which rely on the algebraic structure of RPT mostly studied by Z
and coworkers\cite{Z}\cite{Bre-Ma}\cite{Lo}\cite{La}.
\end{itemize}

The potentialities of EG have however been tested on classes of
popular models -mostly gauge theories, abelian and non abelian- by
G. Scharf (Univ. fo Z\"urich) and coworkers\cite{Scha}\cite{Hu}.

>From the philosophical point of view, EG as well as Z (and K.
Symanzik 1970) formulate RPT in a way that does not require the
use -and removal- of regularizations, in much the same spirit as
was adopted, in concrete cases, to derive properly substracted
dispersion relations\cite{Kae}.

One of the heroic founders of RPT, J. Schwinger\cite{Schwi}, was
obviously attracted by such an approach which he baptized "source
theory" -without realizing that EG had at least settled the
question to all orders of RPT -as a means of curing the trauma
caused by (ultra-violet UV) infinities.

To try and cut a long story short, this program has diffused away
from Z\"urich (where G. Scharf has retired) to Hamburg and
G\"ottingen, under the joined leadership of Michael D\"utsch
(abandoned by Z\"urich) and Klaus Fredenhagen, coworkers and
students\cite{Du-Fre}\cite{Boa}\cite{Bro-Du}\cite{Bre-Du}\cite{Du-Bo}\cite{Pi}.

The road between EG and BPHZ has proved longer than expected.

I will try to summarize some of what has been achieved for RPT on
Minkowski space (which if a very small part of the whole).

\section{Free fields}

It is customary to start from fields whose equations of motion
derive from a Lagrangian.

We will refrain from doing so a priori, because of the famous
example of the free Maxwell field, linear in the creation and
annihilation operators for photons with two helicity states, as
derived form Wigner's representation of the Poincar\'e group for
zero mass, helicity $\pm 1$.

The free Maxwell equations

\begin{equation}
\partial_\mu F_{\mu\nu} = \partial_\mu \tilde{F}_{\mu\nu} = 0
( \tilde{F}_{\mu\nu} = \frac{1}{2} \epsilon_{\mu\nu\rho\sigma}
F_{\rho\sigma}) \label{eq1}
\end{equation}

do not derive form a Lagrangian.

For algebraic reasons which will become manifest as we go along we
shall however land very close to this restricted class of free
fields.

For what concerns us, these are Wightman
fields\cite{Str-Wi}\cite{Wi-Gae}\cite{J}. We shall however give up
the assumption that the Fock space under consideration has a
positive definite Hilbert space metric -allowing for fields which
have proven useful in the framework of gauge theories-. The
connection between spin and statistics can then be jeopardized.

Given a finite set of free fields $\hat{\varphi}$, the ($Z_2$
graded-) commutative algebra $\hat{W}$ of local Wick polynomials
of the field and their derivatives offers a quantum analog for the
space of local interactions\cite{Wi-Gae}.

"As is well known"\cite{Du-Bo},

\begin{equation}
\hat{\mathcal{W}} \sim \mathcal{P}/J(\mathcal{E}) \label{eq2}
\end{equation}

where $\mathcal{P}$ is the algebra of local polynomials of
similarly labelled classical fields and $J(\mathcal{E})$ the ideal
generated by the equations of motion fulfilled by $\hat{\varphi}$.

This "well known" fact has however to be taken with a grain of
salt because, if $\hat{\varphi}$ is to take values in a
representation space of $SL2\mathcal{C} \times SL2 \mathcal{C}$-
with regard to Lorentz covariance- or some "internal" compact
global symmetry group $G$, $\mathcal{P}$ itself comes as a
quotient of $\mathcal{P}^\bigoplus$ (free algebra generated by
monomials $\equiv$ [EG]'s supermultiquadriindices) by an ideal of
relations fulfilled by monomials
\begin{equation}
\mathcal{J}(SL2\mathcal{C} \times SL2 \mathcal{C}) \ \ \ (resp \
\mathcal{J}(G)). \label{eq3}
\end{equation}

For instance $\mathcal{J}(SL2\mathcal{C} \times SL2 \mathcal{C})$
is generated by the relations which express the linear dependence
of three vectors in 2-dimensional
space\cite{Fu-Ha}\cite{Go-Wa}\cite{We}\cite{Pr}
\begin{equation}
0 = \upsilon_1 \wedge \upsilon_2 \wedge \upsilon_3 = \upsilon_1 (
\upsilon_2  \upsilon_3) + \upsilon_2 ( \upsilon_3  \upsilon_1) +
\upsilon_3 ( \upsilon_1  \upsilon_2) \label{eq4}
\end{equation}

where $(\upsilon_i  \upsilon_j) = \upsilon^\alpha_i \ \
\varepsilon_{\alpha\beta}\ \ \upsilon^\beta_j$, $\ \ $
$\varepsilon_{\alpha\beta} = - \varepsilon_{\beta\alpha}, \
\varepsilon_{12} = +1.$

This caveat will take its strength from linearity, resp.
multilinearity, requirements we shall be inclined to enforce on
the following constructions.

Regarding the quotient by $\mathcal{J}(\mathcal{E}) (resp \
\mathcal{J}^\bigoplus(\mathcal{E}))$, we shall only consider the
simplest situation where
\begin{eqnarray}
   \mathcal{P} &=& [\mathcal{W}] \bigoplus \mathcal{J}(\mathcal{E}) \nonumber \\
  resp \ \ \mathcal{P}^\bigoplus &=& [\mathcal{W}]^\bigoplus \bigoplus \mathcal{J}^\bigoplus(\mathcal{E})
  \label{eq5}
\end{eqnarray}

which is almost as strong as requiring that the fields derive from
a non degenerate Lagrangian. Fields fulfilling such a strong
requirement will be baptized regular fields.

Fields which are solutions of a hyperbolic system for which there
are unconstrained Cauchy data on fixed time hypersurfaces are
regular.

This is as broad as we could find a substitute for the usual
canonical formalism.

(N.B.: Restriction at fixed time is legal for distribution
solution of a hyperbolic equation).

In view of these delicacies we have to leave the operator
framework for a functional set up where\cite{Bro-Du}

\begin{eqnarray}
  \hat{\varphi} &\rightarrow & \varphi  \ \ \ \mbox{similarly labelled classical field} \nonumber \\
  :\hat{m}^\alpha(\hat{\varphi}, D\hat{\varphi}):  &\rightarrow &  m^\alpha(\varphi, D \varphi) \ \ \ \mbox{classical monomial} \nonumber\\
  :\ :\hat{m}^{\alpha_1}:(\hat{\varphi}, D\hat{\varphi}) (x_1)\ldots  : \hat{m}^{\alpha_n}:(\varphi, D\varphi)(x_n) :  &\rightarrow &  m^{\alpha_1}(\varphi,
   D\varphi)
   (x_1)\ldots m^{\alpha_n}(\varphi, D\varphi)  (x_n) \ \ \  \nonumber\\
    \mbox{operator product} & \ & \  \mbox{ordinary product} \nonumber\\
  : \hat{F} (\tilde{\varphi}) : : \tilde{G}(\tilde{\varphi}):\ &\rightarrow &  F(\varphi) \ast G(\varphi) \nonumber\\
  \  &=& F(\varphi) exp \left[ i \hbar \int{dxdy} \frac{\overleftarrow{\delta}}{\delta \varphi (x)} \Delta^+(x-y) \frac{\overrightarrow{\delta}}
  {\delta\varphi}(y)\right]  G(\varphi)  \nonumber\\
  (\Omega, \hat{\varphi}(x) \hat{\varphi}(y)\Omega) &=& i \hbar \Delta^+(x-y)\nonumber\\
  (\Omega, \hat{F}(\hat{\varphi})\Omega) &\Rightarrow & \langle F(\varphi)
  \rangle= F(\varphi) \mid_{\varphi = 0} \nonumber\\
  \Omega : \mbox{vacuum state in Fock space} & \ & \
 \label{eq6}
 \end{eqnarray}

$\varphi$  will be taken among smooth functions. Its growth
properties become important in the discussion of the so called
adiabatic limit which will not be touched upon here.

$F$ will be taken form the space of functionals of $\varphi$.

Functionals with arguments from $\mathcal{P}^\bigoplus$ read:

\begin{equation}
F = \sum_n \int dx_1 \ldots dx_n \ F_{\alpha_1 \ldots {\alpha_n}}
(x_1, \ldots x_n) m^{\alpha_1}(\varphi)(x_1) \ldots
m^{\alpha_n}(\varphi) (x_n) \label{eq7}
\end{equation}

where the $F_{[\alpha]}$'s are distribution kernels.

\section{EG's causality condition}

In view of the algebraic delicacies we have mentioned, we shall
work within $\mathcal{P}^\bigoplus$ and functionals thereof. The
effect of interactions is described by a scattering operator in
Fock space. The corresponding functional will be constructed as a
formal power series in a set of smooth coupling functions $\{
g_\alpha \}$ associated with a monomial basis $\{ m^\alpha
(\varphi, D\varphi) \}$ of $\mathcal{P}^\bigoplus$.

Following tradition, we shall write
\begin{eqnarray}
  S(g,\varphi) &=& \sum^\infty_{n=0} S_n (g,\varphi) \nonumber\\
  \  &=& 1 + \frac{i}{\hbar} \int_{M_\tau} d^4 x \sum_\alpha g_\alpha(x) m^\alpha(\varphi,D\varphi)(x) \nonumber\\
  \  &+& \sum_{h\geqslant 2} \left( \frac{i}{\hbar}\right)^n \frac{1}{n!}\int_{M_4^{\times n}} dx_1 \ldots dx_n g_{\alpha_1} (x_1) \ldots g_{\alpha_n}(x_n) \nonumber\\
  \  &\cdot & T \left( m^{\alpha_i} (\varphi, D\varphi)(x_1) \ldots m^{\alpha_n} (\varphi, D
  \varphi)(x_n)\right)
\label{eq8}
\end{eqnarray}

the coefficients of which will be determined recursively.

By construction, they are symmetric in their arguments (which
reflects the commutativity of space time $M_4$, here, Minkowski
space).

The $g$'s are chosen with compact support or in $\mathcal{S}$.
This is partly a technical convenience, some of the physical
content of the construction being concerned with the limit
$g_\alpha (x) \rightarrow g_\alpha$ (cst) for some $\alpha$'s,
(the so called adiabatic limit).

In order to conform with usage, we have kept $\hbar$ ("Planck's
constant") as a formal variable to which are attached some
combinatorial properties of the construction.

The causality requirement is
\begin{equation}
S(g_1 +g_2, \varphi) = S(g_1,\varphi) \ast S(g_2,\varphi)
\label{eq9}
\end{equation}
for $suppg_1, \gtrsim suppg_2$

$\equiv suppg_1 \bigcap suppg_2 + \bar{V}_- = \O)$ where
$\bar{V}^-$ is the closed past light cone). This is called causal
factorization.

\begin{figure}[ph]
\centerline{\psfig{file=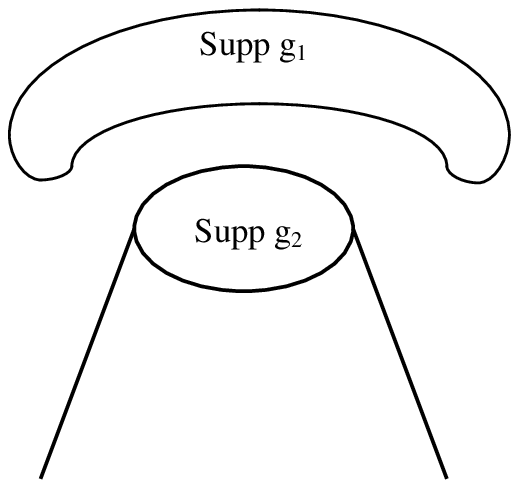,width=1.7in}}
 \vspace*{8pt}
\end{figure}

This is turned by EG into the double recursion hypothesis:

\begin{eqnarray}
  1) \ T^{\alpha_{[n]}} (X_n) &=& T^{\alpha_{[I]}} (X_I) \ast  T^{\alpha_{[I']}} (X_{I'}) \nonumber \\
  \  & X_I \gtrsim X_{I'}& \   \nonumber \\
  2) \ \left[ T^{\alpha_{[n]}} (X_n) \ast T^{\alpha_{[n']}} (Y_{n'}) \right] & = & 0 \nonumber \\
    & X_n \sim Y_{n'} & \nonumber \\
     & n <N , \ n' < N &
     \label{eq10}
\end{eqnarray}
[Notation:
\begin{eqnarray}
\nonumber
 (1, \ldots n) &=& [n] \nonumber \\
  X_n &=& (x_1, \ldots x_n) \ ; \ x_i \in M_4 \nonumber \\
  \alpha_{[n]}  &=& (\alpha_1 \ldots \alpha_n)
  \label{eq11}
\end{eqnarray}

\begin{eqnarray}
  I \subset [n], \ \ I'\subset[n], &\ & I \cup I' = [n] \ \ \ I \neq \varnothing \ \ I' \neq  \varnothing \ \ I \cap I' = \varnothing \nonumber \\
  X_I \gtrsim X_{I'} &\equiv & \left\{ x_i \gtrsim x_{i'} \ i \in I \ i' \in I' \right\} \nonumber \\
  X_n \thicksim Y_{n'}  &\equiv& \left\{ X_n \gtrsim Y_{n'} \ \mbox{and}  \ Y_{n'} \gtrsim X_n
  \right\}.
  \label{eq12}
\end{eqnarray}

At order $N$, one constructs
\begin{equation}
T_I^{\alpha_{[N]}} (X_N) = T^{\alpha_{[I]}} (X_I) \ast
T^{\alpha_{[I']}} (X_{I'}), \ {I \cup I' = [ N ]} \label{eq13}
\end{equation}

whose existence is guaranteed by E.G.'s th0 (which states that
Wick polynomials of free fields can be multiplied by translation
invariant distributions).

Using (1) and (2), one proves
\begin{equation}
T_I^{\alpha_{[N]}} (X_N) = T^{\alpha_{[N]}}_J (X_N) \ \mbox{in}
C_I \cap C_J \label{eq14}
\end{equation}
where
\begin{equation}
C_I = \left\{ X_I \gtrsim X_{I'} \right\}. \label{eq15}
\end{equation}
>From geometry
\begin{equation}
\bigcup_I \ C_I = M_4^{\times N} \backslash D_N \label{eq16}
\end{equation}
where $D_N$ is the diagonal $\{ x_1 = \ldots = x_n \}$.

Using (K. Fredenhagen) a partition of unity $\{\alpha_I\}$
subordinated to the covering $\{ C_I \}$ of $M_4^{\times N}
\backslash D_N$, one defines there\cite{Schwa}
\begin{equation}
\widetilde{T}^{\alpha_{[N]}} (X_N) = \sum_I \ \alpha_I (X_N) \
T_I^{\alpha_{[N]}} (X_N)\label{eq17}
\end{equation}
which, by 1) and 2), is shown to be independent of the choice of
$\{\alpha_I \}$.

Renormalization consists of extending
$\widetilde{T}^{\alpha_{[N]}} (X_N)$ to all of $M_4^{\times N}$
\cite{Ma}\cite{To}.

This involves several steps which constitute the hard core of EG.

\begin{romanlist}[(iii)]
\item[(i)] reducing to scalars : one looks for solutions which
fulfill the Wick Taylor expansion formula:
\begin{eqnarray}
T \left( m^{\alpha_1} (\varphi, D\varphi) (x_1) \right. & \ldots &
\left. m^{\alpha_n}
(\varphi, D\varphi) (x_n) \right) = \nonumber \\
\sum_{\beta\cup\gamma =\alpha} &\ & \langle  T \ m^{\beta_1}
(\varphi,
D\varphi) (x_1)  \ldots  m^{\beta_n} (\varphi, D\varphi) (x_n) \rangle \nonumber \\
\ & \ & \times \ m^{\gamma_1} (\varphi, D\varphi) (x_1) \ldots
m^{\gamma_n} (\varphi, D\varphi) (x_n) \label{eq18}
\end{eqnarray}

\item[(ii)] Reduce $T$ to $T^c$ ($c$ for "connected" through a
"log" algorithm)

\item[(iii)] Enforce translation invariance, by construction.

\item[(iv)] Show that $\langle T^c \rangle$ can be extended (one
way is to use regularizations and renormalization).

\item[(v)] Classify the ambiguity of the extensions. They have
support $D_N$. In the operator formalism there is a theorem
(\cite{EG}, \cite{ES}) which guarantees that it is of the form
\begin{equation}\label{eq19}
\Delta T^{\alpha_{[N]}} (X) = \sum_\beta \ P_\beta^{\alpha_{[N]}}
(\partial) \delta (x_1 - x_N) \ldots \delta (x_{N-1}\ldots x_N)
m^\beta (\varphi, D \varphi) (x_N)
\end{equation}
where $P(\partial)$ is a differential operator with constant
coefficients.

In the off shell formalism, one decides to restrict oneself to
such ambiguities.

\item[(vi)] Power counting theory.

One can restrict the ambiguities so that
\begin{equation}\label{eq20}
deg \ P_\beta^{\alpha_{[N]}} \geq \left[ \sum^N_{i=1}
(\omega^{\alpha_i} -4 ) \right] - (\omega^{\beta} - 4 )
\end{equation}
where the power counting index
\begin{equation}\label{eq21}
\omega^\alpha = \omega (m^\alpha) = \omega \left( \Pi_i
D^{\alpha_i} \varphi^i \right) = \sum_i \left[ \omega(\varphi^i) +
|\alpha_i| \right]
\end{equation}
$\omega(\varphi^i)$ is computable form $\Delta^{+ij}$ $\omega^i +
\omega^j -4 =$ naive scaling dimension of
$\widetilde{\Delta}^{+ij} (p)$.

\end{romanlist}

Remark: (D\"utsch Fredenhagen\cite{Du-Fre}) if one imposes the
Wick Taylor expansion formula to hold then the ambiguity $\Delta
T^{\alpha_N}$ has the above form.

This summarizes a very small (although already quite sizeable!)
part of EG.

\section{Further constraints}

We have already seen one constraint one may wish to impose on  $T$
products (besides symmetry, translation covariance). One needs
more before the space of ambiguities reaches a manageable size,
but one has to be aware of the fact that the constraints one may
wish to impose have to be shown compatible.

Among those which seem to be part of the game and have not found
so far any replacement, are the following.

\subsection{Multilinearity (K. Fredenhagen)\protect \footnote{K. Fredenhagen. Many of the notions used here, besides
multilinearity, are due to him, in writing or otherwise: \protect
\begin{itemize}
    \protect \item identifying the product of functionals as a $\ast$
    product,
    \protect \item identifying AWI as a sufficient condition for the main
    theorem of renormalization to produce an unambiguous answer,
    \protect \item identifying the Wick Taylor expansion formula as the
    solution of a Ward identity.
\protect \end{itemize}}}

 The $T$'s are multilinear in their arguments (e.g., relating $T 2 m^\alpha \ldots $ with
$T(m^\alpha \ldots)$. This seems to be an "obvious" requirement to
make, but it is absolutely not innocent.  It is in particular this
requirement which has lead us to go off shell (and even to
$\mathcal{P}^\bigoplus$).

Going back to the operator formalism in Fock space (D\"utsch
Boas)\cite{Du-Bo} then requires showing that one can construct $T$
products in the functional formalism which belong to
$\mathcal{F}\mathcal{J}(\mathcal{E})$ the $\ast$ ideal in the
$\ast$ algebra of functionals) whenever one argument belongs to
$\mathcal{J}(\mathcal{E})$.

One can do this in the particular case where one can write
\begin{equation}\label{22}
\mathcal{P}^\oplus = \mathcal{J}^\oplus (\mathcal{E}) \oplus
[\mathcal{W}]
\end{equation}
for some representative $[W]$ of $W$, in particular, in the case
of "regular" fields (introduced for this purpose).

On the other hand we have at the moment nothing to say about the
quotient by $\mathcal{J}(SL2 \mathcal{C} \times SL2 \mathcal{C})$
when Lorentz covariance is required.

\subsection{The Action Ward Identity (AWI)}

\begin{equation}\label{23}
\partial^x_\mu T (m^\alpha (\varphi D\varphi(x)\ldots) = T \partial^x_\mu m^\alpha (\varphi , D\varphi)(x)\ldots
\end{equation}

This has several names within BPHZ : solving the routing problem
(W. Zimmermann), energy momentum conservation at each vertex of a
renormalized Feynman Graph), $S(g)$ only depends on $S^1(g)$ not
on the Lagrangian density ...

It can be imposed in the functional formalism (D\"utsch
Fredenhagen\cite{Du-Fre}), much less so in the operator formalism,
the problem there being to find what subset of AWI is compatible
with the quotient by $\mathcal{J}(\mathcal{E})$.

A representative $\mathcal{P}^\oplus_{bal}$ of
$\mathcal{P}^\oplus\ \backslash Pol_+
(\partial)\mathcal{P}^\oplus$ (polynomials without term constants)
can be found, e.g. by going to Fourier transform, for each
monomial and perform the change of variables $(p_1 \ldots p_n
\rightarrow p_1 + \ldots p_n, p_1, \ldots p_{n-1})$ for a monomial
involving $n \neq$ fields and, for identical fields, express
symmetric polynomials in $p_1 \ldots p_n$ in terms of the
symmetric functions $(\sum p_i, \sum_{i \neq j} p_i \otimes p_j +
p_j \otimes p_i, \ldots)$.

Then one can prove
\begin{equation}\label{eq24}
\mathcal{P}^\oplus = \mathcal{P}_{bal} \oplus Pol_+ (\partial)
\mathcal{P}_{bal}.
\end{equation}

Take then an arbitrary solution for $T$, restrict it to arguments
from $\mathcal{P}_{bal}$, and define it on $Pol_+ (\partial)
\mathcal{P}_{bal}$ by using AWI. Check this is a solution, which
fulfills AWI, by construction. This has a very desirable
consequence (D\"utsch Fredenhagen\cite{Du-Fre}): in order to pass
from one solution $S^I(g,\varphi)$, to another
$S^{II}(g,\varphi)$, one can recursively absorb the ambiguities by
which they differ at each order, into counterterms: $S^1 (g,
\varphi) \rightarrow S^1 (g, \varphi) + (\Delta^{I, II}
S^1(g,\varphi )$. This operation is in general not unique (due to
the possibility to perform partial integrations).

It does become unique if the $T$'s are restricted by AWI.

Then the corresponding $S$'s can be written as functionals of
$g_{bal}$ and $\varphi$ (and does not depend on the choice of
$g_{bal})$ and the ambiguity which allow one to go from $S^I$ to
$S^{II}$ acquires a natural group structure (the Stueckelberg
Peterman renormalization group. M. D\"utsch, K.
Fredenhagen\cite{Du-Fre})
\begin{equation}\label{25}
    S^I \left( G_{bal}^{I \ II} (g_{bal}), \varphi \right) =
    S^{II} \left( g_{bal} , \varphi \right)
\end{equation}
where $G^{I \ II}$ is a formal power series, local in $g_{bal})$.

One has
\begin{eqnarray}
  G_{bal}^{I \ III} &=& G_{bal}^{I \ II} \circ G_{bal}^{II \ III}\nonumber \\
  G_{bal}^{I \ II} \circ G^{II \ I} (g_{bal}) &=& g_{bal}.
  \label{26}
\end{eqnarray}

Where $\circ$ is the composition of formal power series.

N.B.: Power counting restrictions are essential for this to make
sense (cf. Bourbaki Alg. Ch IV \cite{Bou}).

The recursively defined ambiguities necessary at each order to
have $S^I$ match with $S^{II}$, collected into one formal power
series local in  $g_{bal}, \Delta^{I \ II}_{g_{bal}} (=
O(g^2_{bal}))$ provide a parametrization
\begin{equation}\label{27}
G^{I \ II}_{bal} (g_{bal}) = g_{bal} + \Delta^{I \ II}_{bal} \circ
G^{I \ II}_{bal}(g_{bal}).
\end{equation}

This is N. Bogoliubov's recursion relation\cite{Bo-Sh}. It is
solved by W. Zimmermann's forest formula\cite{Z} (cf. FM.
Boas\cite{Boa})

N.B.: This is an equation of the type $\underline{y} =
\underline{x} + f (\underline{y})$ which, since Lagrange and
Laplace has prompted a vast amount of literature [e.g. M. Haiman,
W. Schmitt, Jour. Combin. Th. {\bf 50}, 172-185
(1989)\cite{Ha-Sch}. [Thanks to S. Lazzarini for this reference].

The particular case of RPT has been closely scrutinized  by A.
Connes, D. Kreimer)\cite{Co-Kr}.

Actually AWI is not only sufficient but necessary if one wants the
ambiguities to be endowned with a group structure.

This is however not yet the renormalization group of BPHZ for
which one has to reduce $\mathcal{P}^{(\oplus)}$ to
$\mathcal{W}^\oplus$ and $g_{bal}$ to $g_{bal}^{phys}$ (the
"physical" coupling constants to be defined).

This has been done in the case of regular fields.

\subsection{The Wick Taylor expansion formula}

already mentioned, optional.

\subsection{Connectedness and the $\hbar$ expansion \protect \footnote{The combinatorics of the connected can be done either as
in \protect \cite{EGS} or using Ruelle's $\ast$ product \protect
\cite{Ru} which is the dual of the commutative coproduct standard
for tensor algebras \protect \cite{Bou}.}}

Connectedness has also has been used in the construction of a
solution.

A combinatorial property of the Wick Taylor expansion formula is
that the connected $T^c$'s are formal power series in
$\sqrt{\hbar}$ - and their vacuum expectation values formal power
series in $\hbar$- at the heuristic unrenormalized Feynman graph
level or at the level of the Wightman functions.

One may impose this as M. D\"utsch, K. Fredenhagen, F. Brennecke
do. Or, one may derive \cite{Bru-Fre} it from a naturalness
assumption according to which, within equivalence classes of free
fields, isomorphisms should give rise to one to one
correspondences between the corresponding $T$ products. Here,
apply this to $\varphi$ and $\varphi \sqrt{\hbar}$, and $\varphi
\rightarrow -\varphi$.

Combined with multilinearity, this puts constraints on ambiguities
allowed by power counting.

Naturalness also applies to Lorentz covariance and covariance
under compact internal symmetry groups which may be parts of the
attributes of the free fields.

\section{General properties of R.P.T.}

\subsection{Local insertions and the renormalized action principle}

Much of the combinatorial structure of RPT is connected with the
renormalization group structure of the ambiguities.

The corresponding Lie algebra is the Lie algebra of "local
insertions"
\begin{equation}\label{28}
    \Delta = \int d^g x \ \sum_\alpha \ \Delta_\alpha (g,Dg)(x) \frac{\delta}{\delta g_\alpha(x)}
\end{equation}
(where $g$ means $g_{bal}$), where the $\Delta_\alpha$'s are local
and constrained by power counting.

They are compatible with the causal factorization property
\begin{equation}\label{29}
    (\Delta S)(g_1+g_2,\varphi) = (\Delta S)(g_1,\varphi) \ast
    S(g_2, \varphi) + S(g_1, \varphi)\ast (\Delta S)(g_2,\varphi)
\end{equation}
$supp g_1 \gtrsim supp g_2$, as a consequence of the locality of
$\Delta_\alpha$ and the commutativity of $\Delta$ with the $\ast$
product.

As a result, $S+\Delta S$ fulfills causal factorization up to
$O(\Delta^2)$.

Any $\Delta$ with this property therefore provides a $\Delta S$
which is an infinitesimal ambiguity, and therefore has the above
form.

The totality of such $\Delta$'s (local derivations of the $\ast$
algebra of functionals) is not known. Some particular cases give
rise to the so called "renormalized action principles"
(Lowenstein\cite{Lo}, Lam\cite{La}, Breitenlohner,
Maison\cite{Bre-Ma}, D\"utsch\cite{Bre-Du}). For instance:

\begin{eqnarray}
  \Delta_\varphi &=& \int \ d^4x \ \Delta^i (g,Dg)(x) \ \frac{\delta}{\delta \varphi^i(x)} \nonumber \\
  \Delta_{t \varphi} &=& \int \ d^4x \ \Delta (g,Dg)(t\varphi)^i \ \frac{\delta}{\delta \varphi^i(x)} \ \ \mbox{if} \ [t, \Delta^+] = 0\nonumber \\
  \Delta_{\mathcal{E} \varphi} &=& \int \ d^4x \ \Delta (g,Dg)(x) \ \mathcal{E}(\partial) \ \varphi(x) \ \frac{\delta}{\delta \varphi(x)} \ \
  \mbox{if}\  \mathcal{E}(\partial) \Delta^+ = 0
\label{30}
\end{eqnarray}

The RAP's are usually presented within the framework of the
Lagrangian formalism, for functionals introduced in that
framework, in the adiabatic limit which we have not touched upon.
Here, we shall limit ourselves to what we think is a key step in
this direction, namely isolating the role of the free field
equations of motion.

Assume one can find $\mathcal{P}_{bal}$ such that
\begin{eqnarray}
  \mathcal{P}_{bal} &=& \mathcal{P}_{phys} \oplus \mathcal{P}_{\mathcal{J}(\mathcal{E})} \ \ (\mbox{all elements} \in \mathcal{J}(\mathcal{E})) \nonumber \\
g_{bal} &=& \left( g_{phys} , g_\mathcal{E} \right) \label{31}
\end{eqnarray}

This can be done (with some efforts for regular ($\supset$
Lagrangian) fields. One can find $\sum_c(g_{phys},g_\mathcal{E},
\varphi)$ such that
\begin{equation}\label{32}
    \frac{\delta\Sigma_c}{\delta g_\mathcal{E}} \subset
    \mathcal{F}\mathcal{J} (\mathcal{E})
\end{equation}
the ideal generated by $\mathcal{J}(\mathcal{E})$ in the space of
functionals.

This can be proved by recursion. Then, one has
\begin{equation}\label{33}
    S^c \left( g_{phys}, g_{\mathcal{E}}, \varphi \right) = \Sigma_c \left( G_{phys}(g_{phys} , g_\mathcal{E}) \ G_{\mathcal{E}} (g_{phys}, g_\mathcal{E}), \varphi
    \right)
\end{equation}

Differentiating with respect to $g_\mathcal{E}$ and taking vacuum
expectation values, which annihilates terms belonging to
$\mathcal{F}\mathcal{J}(\mathcal{E})$, one gets
\begin{equation}\label{34}
    \left\langle \frac{\delta S^c}{\delta
    g_\mathcal{E}}\right\rangle= \frac{\delta G_{phys}}{\delta
    g_\mathcal{E}} \left( \frac{\delta G_{phys}}{\delta g_{phys}}
    \right)^{-1} \left\langle \frac{\delta S_c}{\delta
    g_{phys}}\right\rangle
\end{equation}

In other words, equations of motion multiplied by composite
operators are expressible in terms of physical couplings. This
contains RAP for the usual $\mathcal{Z}^c (g,J)$. (see section
5.3).

\subsection{Models}

Models are defined by submanifolds in the space of coupling
functions which are stable under the action of the renormalization
group.

This is an old idea (e.g., scalar electrodynamics has one more
parameter than spinor QED, namely a quartic self coupling of the
scalar field). This has been revived under the name "reduction of
coupling constants" by W. Zimmermann, R. Oehme, K. Sibold and
followers\cite{Z}.

Most models are defined by a system of Ward identities in
involution -modulo the proof that they can be fulfilled "without
anomalies"-, besides AWI. Ex. $g_\alpha =0$ $\omega^\alpha >4$
(renormalized models).

\subsection{Contact with the conventional functional formalism}

Separating $g$ and $j$. the coupling function of the field itself
one can show that, given a $\Delta_F$,
\begin{equation}\label{35}
    S \left( g, j=0 ; \varphi + \Delta_F \ast j \right)
    {\mbox{\large{e}}}^{\frac{i}{\hbar}\langle j,\varphi \rangle +
    \frac{i}{2\hbar}\langle j, \Delta_F \ast j \rangle }
\end{equation}
solves the causality program.

If the field derives from a Lagrangian $\left( \Delta^{\ast-1}_F =
Pol(\partial) \delta = K \delta \right)$, the $\ast$ product is
expressible in terms of $j$. For $\varphi = 0$
\begin{equation}\label{36}
    Z ( g, j) = S \left( g, j=0 , \Delta_F \ast j \right)\ {\mbox{\large{e}}}^{\frac{i}{2 \hbar}\langle j, \Delta_F^\ast  j \rangle }
\end{equation}
fulfills causality with respect to $g,j$ with
\begin{equation}\label{37}
    \ast = exp \frac{i}{\hbar} \int dx dy \frac{\delta}{\delta
    \overleftarrow{j}(x)}\overleftarrow{K}_x \Delta^+ (x-y)
    \overrightarrow{K}_y
    \frac{\delta}{\delta\overrightarrow{j}(y)}.
\end{equation}

One then defines the IPI generator $\Gamma (g,\varphi)$ by
Legendre transform.

This can be done starting from $S^c (g,o;\varphi)$ itself without
the need of a Lagangian:
\begin{equation}\label{38}
    S^c (g,o;\varphi) = \Gamma'(g,\phi) - \frac{1}{2}
    \left( \frac{\delta\Gamma'}{\delta\phi'}, \Delta_F \ast
    \frac{\delta\Gamma'}{\delta\phi}\right) |_{ \phi = \varphi + \Delta_F
    \ast \frac{\delta\Gamma'}{\delta\phi}}
\end{equation}
which is inverted by
\begin{equation}\label{39}
    \Gamma' (g,\phi)=S^c(g,0;\varphi) +\frac{1}{2} \left( \frac{\delta S^c}{\delta
    \varphi}, \Delta_F \ast \frac{\delta S^c}{\delta \varphi} \right)
    |_{\varphi = \phi - \Delta_F \ast \frac{\delta S^c}{\delta
    \varphi}}
\end{equation}
($\Gamma'$ is the interacting part of $\Gamma$, i.e. the
"effective interaction").

It is customary, in view of the adiabatic limit to eliminate from
both $Z$ and $\Gamma$ the term independent of $j$, resp.
$\varphi$, and to define $\Gamma$, as well as $\Gamma'$ in such a
way that it starts with terms quadratic in $\phi$. $Z^c$ has a
term linear in $j$ with coefficient  $F=\Delta_F \ast \frac{\delta
S^c}{\delta \varphi}|_{\varphi=0}$ which can be absorbed in the
Legendre transform formula: it suffices to change $\phi$ into
$\phi+F$ in the stationarity condition in order to have a
$\Gamma'$ which starts quadratically in $\phi$.

\section{Conclusion and outlook}

There are still many "details" to be filled in, and, if possible,
simplified in comparison with the existing proofs. There are also
some "terrae incognitae". Here are some, belonging to either
species.

\subsection{$\mathcal{J}(\mathcal{E})$}

Besides regular fields which may be slightly more general than
those deriving from a non degenerate Lagrangian, the only case
which has been looked at is that of the Maxwell field
$F_{\mu\nu}$, which provides some understanding of the collection
of exotic fields used in the perturbative treatment of gauge
theories. This has been started by Michel Dubois-Violette who
found a geometrical characterization of the Faddeev Popov ghost
(mostly unpublished because of difficulties with the tip of the
light cone). This has been continued (R.S. in "Fifty years of Yang
Mills" 2004) but is by no means complete.

\subsection{Extending distributions}

It may be worthwhile studying $\langle \widetilde{T}^\lambda (x)
\rangle = \langle \widetilde{T}^(\lambda x) \rangle$ for $\lambda
>0$ and its Mellin transform (cf. M. Berg\'ere and YMP Lam circa
1975). This may provide a substitute for the dimensional complex
parameter $\epsilon$, without the group theoretical drawbacks
associated with

\subsection{$\mathcal{J}(SL2\mathbb{C} \times SL2 \mathbb{C})$}

\subsection{Connexity and 1PI}

There is some nice combinatorics associated with connexity (F.
Patras, M. Schocker 2005\cite{Pa-Sch}, \cite{Eb-Pa}). It would be
nice to have a direct algebraic proof for the connexity of the
Ruelle Araki products (EGS 75)\cite{EGS} and a streamlining of the
proof of the corresponding spectral properties. Same for 1PI.

\subsection{The adiabatic limit}

If the distributions $\langle \widetilde{T}(X)\rangle , \langle
T(X) \rangle $ are defined as temperate, (the coupling functions
belonging to $\mathcal{S}$, it may be desirable to study $\langle
\widetilde{T}^\lambda (X)\rangle$ for $\lambda \rightarrow \infty$
by lifting them to a suitable compactification of $M_4^{\times
|X|}$ and view the adiabatic limit as an extension problem on this
compactification, to the compactification set (manifold).

The recursion procedure together with a characterization of the
ambiguities may lead to an infrared renormalization group.

\subsection{From the off shell (functional) framework to the on shell (operator) framework}

The connection between the two set ups, which plagues quantum
field theory deserves more care.

\section*{Acknowledgments}

I have drawn much inspiration form G. Barnich, M. Bauer, C.
Becchi, J. Bros, C. Brouder, M. Dubois-Violette (!), M. D\"utsch,
H. Epstein, K. Fredenhagen, G. Girardi, M. Henneaux, T. Hurth, C.
Imbimbo, S. Lazzarini, F. Patras, G. Scharf, Th. Sch\"ucker, K.
Sibold, I. Todorov, F. Thuillier, whom I thank for wasting some of
their time responding to often silly questions.

The points of view expressed here reflect the desire to try and
put together some of the field theory I learnt from R. Haag, R.
Jost, G. K\"allen, D. Kastler, H. Lehmann, B. Schroer, J.
Schwinger, K. Symanzik, A.S. Wightman, W. Zimmermann.

\appendix

\section{More general perturbations}

During the colloquium Manuel Asorey asked the tantalizing
question: can one describe interactions among more general local
fields, beyond free fields. This is not only a natural question:
it has been faced within the study of integrable perturbations of
conformal fields.  A systematic renormalized perturbation theory
does not exist, however. We shall sketch out what seem to be the
hardest obstructions to such a construction.

So let $\widehat{\varphi}$ be some Wightman fields (characterized
by Wightman functions $W (x_1 \ldots x_n) =
(\Omega,\widehat{\varphi} (x_1) \ldots \widehat{\varphi} (x_n)
|0)$.

We need a space of interactions; it is natural to choose the
Borchers class of $\widehat{\varphi}$. If $\widehat{\varphi}$ is
described by some renormalized perturbation theory, the Borchers
class will be labelled by local polynomials in $\widehat{\varphi}$
just as that of some corresponding free fields. The model we have
in mind is described in EG:
\begin{equation}\label{40}
    V(g,h) = S^{-1} (g) \ S(g+h)
\end{equation}
fulfills causal factorization for all $g$'s, with respect to  $h$
if $S(g)$ is perturbatively defined with the latter causal
property. We assume that the adiabatic limit $g(x) \rightarrow g$
(cst) exists. Together with the corresponding field
$\widehat{\varphi}$, we have the local monomials
$\widehat{m}^\alpha (\widehat{\varphi})$.

As in the free field case causal factorization allows to describe
the recursive construction (in powers of $h$) as an extension
problem through the diagonal at the level of functionals.

The construction gets stuck, with the present technology at the
level of the reduction to scalars:

The Wick Taylor formula can be generalized following a
construction due to AS Wightman and J. Challifour (1966
unpublished) modulo a slight generalization to include all
$\widehat{m}^\alpha$'s Wick products $\vdots \ \cdot \ \vdots$ can
be defined by the Wick Taylor formula:
\begin{eqnarray}\label{41}
    \widehat{m}^{\alpha_1} (x_1) \ldots \widehat{m}^{\alpha_n}
    (x_n) &=& \sum_{\beta_i \cup \gamma_i = \alpha_i} \left( \Omega, \widehat{m}^{\beta_1} (x_1) \ldots \widehat{m}^{\beta_n} (x_n) \Omega
    \right) \nonumber \\
    && \vdots \widehat{m}^{\gamma_1} (x_1) \ldots
    \widehat{m}^{\gamma_n}
    (x_n)\vdots
\end{eqnarray}
with the convention $\widehat{m}^\phi =1$, and there follows
Wick's theorem
\begin{eqnarray}\label{42}
    \vdots \  \widehat{m}^{[\alpha]}(X) \ \vdots \ \cdot \ \vdots
    \ \widehat{m}^{[\beta]}(Y)&=& \sum
    \ C^{[\alpha][\beta]; [\alpha'][\beta']}(X \cup
Y) \ \widehat{m}^{[\alpha''] \cup [\beta'']}(X \cup
    Y)\nonumber \\
    && [\alpha_j \cup [\alpha'']] =[\alpha] \nonumber \\
    && [\beta_j \cup [\beta'']] =[\beta]
\end{eqnarray}
where the contraction symbol
$C^{[\alpha][\beta];[\alpha'][\beta']}(X \cup Y)$ is given by
\begin{eqnarray}\label{43}
  C^{[\alpha][\beta];[\alpha'][\beta']} &=& \sum_k \prod_\kappa
  \left\langle \widehat{m}^{[\alpha'_\kappa] [\beta'_\kappa]} (X \cup Y) \right\rangle^T \nonumber \\
   && \bigcup_\kappa [\alpha'_\kappa] = [\alpha'] \nonumber \\
  && \bigcup_\kappa [\beta'_\kappa] = [\beta'] \nonumber \\
  && \alpha'_n \cdot \beta'_\kappa \ \ \mbox{not simultaneously empty}
\end{eqnarray}
$\langle \ \ \rangle^T$ refers to the truncated expectation
values.

The difficulties with these Wick prodcuts $\vdots \cdot \vdots$ is
that, in general
\begin{enumerate}
    \item they only fulfill local commutativity (not, necessarily,
    full commutativity as $\vdots \ \ \vdots$)
    \item they are insufficiently renormalized (only vacuum
    contributors to divergences are substracted out), so that they
    do not fulfill theorem $0$ of EG (multiplicability by
    translation invariant distributions.
\end{enumerate}

Note however that using
\begin{equation}\label{43}
\widehat{m}^{\alpha}(x) = \vdots \ \widehat{m}^{\alpha}(x) \
\vdots
\end{equation}
(assuming $\langle \widehat{m}^{\alpha}(x)\rangle =0$).

We can apply Wick's theorem to deduce
\begin{equation}\label{44}
\langle \widehat{m}^{\alpha_1}(x_1) \ldots
\widehat{m}^{\alpha_n}(x_n) \rangle = C^{\alpha_1 \ldots \alpha_n}
(x_1 \ldots x_n)
\end{equation}
where $C^{\alpha_1 \ldots \alpha_n} (x_1 \ldots x_n)$ is given by
a sum of products of truncated functions which can be associated
with a graph involving not only oriented lines joining two
different vertices but also oriented circles joining larger
subsets of points from ($x_1 \ldots x_n)$.

These expressions are well defined as distributions because of the
spectral properties which imply that the $W^T$'s are boundary
values of functions holomorphic in tubes in the difference
variables, which allows to multiply boundary values.

The problem is now to define the corresponding time ordered
products recursively. In the free field case, as already
mentioned, there are two crucial properties:
\begin{enumerate}
    \item the Wick algebra is commutative
    \item it admits translation invariant distribution
    coefficients (THO)
\end{enumerate}

Concerning 1) one may extend the above combinatorics to $T$
products, which looks somewhat circular since it involves $\vdots
\  T(\cdot) \ \vdots$'s for which there is no natural definition.
2) is even more problematic. In case of emergency, one may try to
renormalize the time ordered versions of the contraction symbols.
The corresponding combinatorics has been studied (excluding
renormalization) in a recent article which the authors kindly sent
me upon return from this conference: C. Brouder, A. Frabetti, F.
Patras: "Decomposition into one particle irreducible Green
functions in many body physics" : arXiv: 08033747, v.1
[cond-mat-str-el], 26 Mar 2008.



\begin{thebibliography}{0}

\bibitem{Bo-Sh} N.N. Bogoliubov, D.V Shirkov, \textit{Introduction to the theory
of quantized fields.} Moscou 1956.

\bibitem{He} K. Hepp, {\it Comm. Math.
Phys.} {\bf 2} (1966), 301, \\
Th\'eorie de la Renormalisation, Lectures Notes in Physics,
Springer (1969)\\
Les Houches (1970)\\
H.P.A. {\bf 36} (1963) 355.

\bibitem{Z} W. Zimmermann, \\
1970 Brandeis Summer Institute Vol. 1 \\
MIT Press (1971) \\
Com. Math. Phys. {\bf 11} (1968), 1, {\bf 15} (1969), 208\\
Com. Math. Phys. {\bf 97} (1985) 211\\
R. Oehme, W. Zimmermann, Com. Math. Phys. {\bf 97} (1985) 569.

\bibitem{Ste} O. Steinmann, Lecture Notes in Physics, Vol. II, Springer 1971.

\bibitem{Wei} S. Weinberg, {\it The Quantum Theory of Fields}, Vol. I, Cambridge University Press (1995).

\bibitem{Stu-Pe} ECG St\"uckelberg, A. Petermann, {\it HPA} {\bf 26} (1953), 499.

\bibitem{Stu-Ri} ECG St\"uckelberg, D. Rivier, {\it HPA} {\bf 23} (1950), 215-22.

\bibitem{EG}
H. Epstein, V. Glaser, {\it Ann. IHP} {\bf A19} (1973), 211, Les
Houches 1970.

\bibitem{Bre-Ma} P. Breitenlohner, D. Maison, {\it Comm. Math.
Phys.} {\bf 52} (1977), 11--38.

\bibitem{Lo} J.H. Lowenstein, {\it Comm. Math.
Phys.} {\bf 24} (1971), 1--21.

\bibitem{La} YMP Lam, , {\it PRD} {\bf 6} (1972), 2145--2161.

\bibitem{Scha} G. Scharf, {\it Finite Quantum Electrodynamics}, Springer 1995, {\it Quantum gauge theories,
a true ghost story}, J. Wiley (2001).

\bibitem{Hu} T. Hurth, {\it Ann. Phys.} {\bf 244} (1995), 340--425\\
T. Hurth, M. Skenderis, {\it Nucl. Phys.} {\bf B541} (1999),
566--614.

\bibitem{Kae} G. K\"allen, {\it Quantenelectrodynamik}, Handbuch der Physik (S. Fl\"ugge Ed.), Bds Teil1, Springer 1958.

\bibitem{Schwi} J. Schwinger, {\it Particles Sources and Fields}, Vol. I, II, III, Addison Wesley 1970-1973.

\bibitem{Du-Fre}
M. D\"utsch, K. Fredenhagen, Rev. Math. Phys. {\bf 16} (2004)
1291-1348, \\
{\it Comm. Math. Phys.} {\bf 243} (2003), 275--314 \\
aeXiv: hep-th/0501228, 28 Jan. 2005.

\bibitem{Boa} F.M. Boas, hep-th/0001014.

\bibitem{Bro-Du} C. Brouder, M. D\"utsch, 9 Oct. 2007.

\bibitem{Bre-Du} F. Brennecke, M. D\"utsch, arXiv, 07053160 (hep-th).

\bibitem{Du-Bo}
M. D\"utsch, F.M. Boas, Rev. Math. Phys. {\bf 14} (2002) 977.

\bibitem{Pi} G. Pinter, hep-th 9911063.

\bibitem{Str-Wi} R.F. Streater, A.S. Wightman, PCT, Spin and Statistics and all that, Benjamin.

\bibitem{Wi-Gae} A.S. Wightman, L. G\"arding, ArXiv f\"ur Physik (1965), 129-184.

\bibitem{J} R. Jost, {\it The general Theory of quantized fields AMS}, vol. IV (1965).

\bibitem{Fu-Ha} W. Fulton, J. Harris, {\it Representation Theory, a first course}, Grad. Texts Math. Springer (1991).

\bibitem{Go-Wa} R. Goodman, N.R. Wallach, {\it Representations and Invariants of the Classical Groups},
[Encyclopedia of Mathematics and its Applications], Cambridge
Univ. Press (1998).

\bibitem{We} H. Weyl, {\it Classical Groups} Princeton Univ. Press (1946).

\bibitem{Pr} C. Procesi, {\it Lie Groups (an approach thrgough invariants and representations)} Springer (2007).

\bibitem{Schwa} L. Schwartz, {\it Th\'eorie des distributions}, Hermann.

\bibitem{Ma} B. Malgrange, {\it Ideals of Differentiable functions}, (Tata Lectures), Oxford Univ. Press (1966).

\bibitem{To} J.C. Tougeron, {\it Id\'eaux de fonctions Diff\'erentiables} (Ergebnisse des Math... Bd71), Springer 1972.

\bibitem{ES}
H. Epstein,  {\it Nuov. Cim. } {\bf 27} (1963), 886 \\
B. Schroer, unpublished.

\bibitem{Bou} Bourbaki, Alg\`{e}bre Ch IV.

\bibitem{Ha-Sch} M. Haiman, W. Schmitt, {\it Journal of Combinatorial Theory}, Series {\bf A50} (1989), 172--185.

\bibitem{Co-Kr} A. Connes and D. Kreimer, CMP {\bf 119} (1998),203; Lett.
Math. Phys {\bf 48} (1999), 85; JHEP {\bf 09} (1999), 24; CMP {\bf
210} (2000), 249; CMP {\bf 216} (2001), 215.

\bibitem{EGS}
H. Epstein, V. Glaser, R. Stora, Les Houches 1975, J. Bros, D.
Iagolnitzer.

\bibitem{Ru} D. Ruelle, {\it Statistical Mechanics}, Benjamin.

\bibitem{Bru-Fre} R. Brunetti, K. Fredenhagen, R. Versch, arXiv: math-phys/01, 12041, 19
Dec. 2001.

\bibitem{Pa-Sch} F. Patras, M. Schocker, Adv. in Math. 2005.

\bibitem{Eb-Pa}
K. Ebrahimi-Fard, F. Patras,\\
arXiv 07105134 (math-phys), 26 Oct. 2007 88 arXiv 0705 [math-Co],
9 May 2007.

\end{thebibliography}
\end{document}